# An alternative aggregate preference ranking algorithm to assess environmental effects on macrobenthic abundance in coastal water

[1]Mohammad Gholizadeh, [2]Majid Zerafat Angiz L*, [3]Seyed Mahmoud Davoodi, [4]Rajab Khalilpour, [1,5] Anita Talib, [1,6]Khairun Yahya, [2]Sahubar Ali Nadhar Khan

1. Centre for Marine and Coastal Studies, Universiti Sains Malaysia, 11800 Penang, Malaysia
2. School of Quantitative Sciences, Universiti Utara Malaysia, 06010 Sintok, Kedah, Malaysia
3. Islamic Azad University, Abadeh Branch, Department of Mathematics, Abadeh, Iran
4. School of Chemical and Biomolecular Engineering, The University of Sydney, Australia
5. School of Distance Education, Universiti Sains Malaysia, 11800 Penang, Malaysia
6. School of Biological Sciences, Universiti Sains Malaysia, 11800 Penang, Malaysia

* Corresponding author. Tel.: 0060173509310;

E-mail address: mzerafat24@yahoo.com

**Abstract**

Coastal marine waters are ranked among the most important aquatic ecosystems on earth in terms of ecological and economic significance. Since the Industrial Revolution, human activities have drastically changed coastal marine ecosystems. The development of rules and regulations to protect these ecosystems against human activities requires availability of environmental assessment standard. This necessitates the identification of the key parameters that reflect condition of the coastal water ecosystem. Macrobenthic assemblages are recognized to rapidly respond to changes in the quality of water or habitat. Therefore, it would be useful to study the population of macrobenthos and assess the influential factors on the growth of this species.

This study is categorized as multidisciplinary approach which contains two perspectives; ecological and mathematical. In the ecological section, the effect of the water quality parameters (e.g. pH, temperature, dissolved oxygen and salinity) and the sediment characteristics on the macrobenthic abundance is studied. A total of 432 samples were collected and analyzed from four touristic costal locations (at various distances form the coast) of Penang National Park to investigate the spatial change of macrobenthic assemblage.

In a mathematical perspective, this paper pursues a new algorithm based on the performance evaluation methods. For this purpose, first, Data Envelopment Analysis (DEA) is employed to evaluate a group of Decision Making Units (DMUs). Consequently, inputs of the mentioned DMUs are considered as alternatives (or candidates), and using a modified DEA model that is categorized as aggregating preference ranking method, the influence of inputs in efficiency of DMUs is investigated.

**Keyword**: macrobenthic assemblage; environmental parameters; Penang Island; Data Envelopment Analysis (DEA), Aggregation Preference Ranking.

## 1. Introduction to macrobenthic community

It has long been recognized how human actions have changed terrestrial environments (Leopold, 1933) and according to the Intergovernmental Panel on Climate Change's (IPCC) fifth assessment report, today "human influence on the climate system is clear" (Ipcc, 2014). It has been also acknowledged that human activities have drastically changed marine ecological system (Jackson et al. 2001). Ecology is the integrated survey of the relation of living organisms comprising human beings to their environment (Cuff and Goudie, 2009, Chapin et al., 2011). The study of living organisms in relation with their abiotic environments is one of the key research fields of ecological science. The climate change during the last few decades has made the study of population, diversity, biomass, and distribution of organisms at the center of ecological research attention. The study of organisms' quantity and quality at any specific location allows scientists to find the organisms' niche living standards. Such studies require assessment of the influential and fundamental factors on the population growth or decline in order to control abundance of a species.

Coastal marine waters rank among the most important aquatic systems on earth in terms of ecological and economic significance (Kennish, 1997). Macrobenthos embraces a group of organisms whose habitat is the bottom of a water body, and consists of variety of species such as

Polychaetes, Pelecypods, Anthozoans, Molluscs, Crustaceans, Echinoderms, etc. The abundance of macrobenthos principally depends on chemical and physical properties of the substratum. Furthermore, macrobenthic assemblages are recognized to respond to changes in the quality of water or habitat. Therefore, it would be useful to study the population of macrobenthos and assess the influential factors on the growth of this species.

Figure 1 illustrates the Eutrophication process within the coastal waters. The discharge of waste water rich in nutrient components such as phosphate and nitrate results in the increase of phytoplanktons and algae in water body. This reduces or blocks oxygen flux to lower levels of water which is addressed as hypoxia. The consequence is the death of some macrobenthos and migration of others to higher levels (closer to shore). Therefore, extended residence period in particular habitats and presence or absence of specific benthic organisms in a particular environment can be utilized as bio-indicators of habitat conditions change over time. Although a variety of organisms are available for the study of subtidal environments, soft-sediment macrobenthos are most commonly utilized and are likely the most appropriate choice due to their lack of mobility. As such their observed variations over time could be attributed to the contamination stress rather than to migration or movement (Clark, 1997).

**Figure 1:** Illustration of Eutrophication (Image: (Hillewaert, 2006))

The major effective factors on the abundance of macrobenthos are categorized into two aspects including water and sediment characteristics. Water characteristics consist of pH, temperature, dissolved oxygen and salinity. On the other hand, medium sand (~425 µm), fine sand (~250 µm), very fine sand (~125 µm), and silt are known as sediment parameters. There are many different ways to relate macrobenthic assemblage structure to water quality with many new analysis methods. According to Hellawell (1986), the macrobenthic assemblage is considered as an indicator of water quality. More detailed characteristics of macrobenthos such as size, the life cycle, responses to variations in natural and human activity conditions, and nutrition are used rather than smallest invertebrates and protozoans for ecological study and modelling. For instance, numerous macrobenthos live in the substratum of sediment (e.g., polychaetes, clams) or on the sediment (e.g., crustacean, crabs). Further, limited movement and response to ambient condition of macrobenthic communities have been the topic of investigations for anthropogenic effect. The abundance and composition of macrobenthic fauna are in a close relation to the water characteristic of aquatic habitats (Costa et al. 1997; Weerasundara et al. 2000; Pathiratne & Weerasundara 2004) and hence, macrobenthic fauna are regarded to be good indicators of the previous and present statuses of water (Cowell et al. 1975, Gamlath & Wijeyaratne 1997).

In summary, macrobenthic assemblages are considered to be good indicators of ecosystem health due to their strong connection with water and sediment qualities (Dauer et al., 2000; Muxika et al., 2007; Bakalem et al., 2009). Therefore, the first step in environmental performance assessment is to identify the key parameters that reflect condition of the coastal water ecosystem. Ecological models are increasingly utilized in investigation and decision-making to enhance the estimation confidence in the model outputs. The desire for rigorous methodologies has resulted in the introduction of various approaches and techniques. Undoubtedly, application of a proper technique depends on the problem nature, the data, and the purposes of modeling. This research aims to study the impact of the parameters solely and interactively. In fact, we would like to rank the above-mentioned parameters in terms of their impacts on the population growth of macrobenthos. This is implemented by using a mathematical algorithm in group decision making circumstances. More precisely, we aim to examine the effects of environmental factors on macrobenthic abundance using DEA with a case-study of the north west of Penang Island.

## 2. Data Envelopment Analysis

Group or collaborative decision-making technics embrace all methods and algorithms that aggregate individuals' viewpoints to make a collective choice. There are numerous approaches in aggregating preferential votes (Hosseinzadeh Lotfi et al., 2013). Among these, Data envelopment analysis (DEA) is a powerful tool for evaluating the performance of organizations and their functional units. DEA assesses the relative efficiency of a set of homogenous decision-making units (DMUs) by using a proportion of the weighted sum of outputs to the weighted sum of inputs. DEA is a non-parametric method, and can be utilized to recognize the source and the amount of ineffectiveness in each input relative to each output for the aimed decision-making units.

In mathematical terms, consider a set of $n$ DMUs, in which $x_{ij}(i=1,2,...,m)$ and $y_{rj}(r=1,2,...,s)$ are input and output of $DMU_j$ ($j=1, 2,...,n$). A DEA model for assessing $DMU_p$ which is known as the MAJ model (Saati et al. 2001) is formulated as,

$$w_p^* = \min w + 1 \tag{1}$$
$$s.t.$$
$$\sum_{j=1}^{n} \lambda_j x_{ij} \leq x_{ip} + w$$
$$\sum_{j=1}^{n} \lambda_j y_{rj} \geq y_{rp} - w$$
$$\lambda_j \geq 0 \qquad j=1,2,...,n$$
$$w \text{ free}$$

In Model (1), the optimal value ($w_p^*$) demonstrates the relative efficiency score associated with $DMU_p$ which is under evaluation. $\lambda_j$ is a nonnegative value related to the $j^{th}$ DMU. The vector $\lambda = (\lambda_1, \lambda_2,..., \lambda_n)^t$ is used to construct a hull that covers all of the data points. In this model, $DMU_p$ is efficient if $w_p^* = 1$. Basically, the model finds the input and output weights that will provide the best possible efficiency score to $DMU_p$.

Preference aggregation problem, in the context of a ranked voting system is a group decision making problem of selecting $m$ alternatives from a set of $n$ alternatives ($n > m$). Hence, each decision maker ranks the alternatives from the most preferred (rank = 1) to the least preferred (rank = $n$). Obviously, due to different opinions of the decision makers, each alternative may be placed in a different ranking position. Some studies suggest a simple aggregation method by finding the total score of each alternative as the weighted sum of the votes received by each alternative according to different decision makers. In this method, the best alternative is the one with the largest total score. The key issue of the preference aggregation is how to determine the weights associated with different ranking positions. Perhaps, Borda–Kendall method (Hashimoto 1997) is the most commonly used approach for determining the weights due to its computational simplicity.

Cook and Kress (1990) proposed a method that is based on DEA to aggregate the votes from a preferential ballot. For this purpose, they used the following DEA model (2) in which outputs are the number of first place votes, second place votes and so on, that a DMU has obtained.

$$\max \ \beta_p = \sum_{k=1}^{n} \mu_k \omega_{pk} \qquad (2)$$

s.t.

$$\sum_{k=1}^{n} \mu_k \omega_{jk} \leq 1 \qquad j = 1, 2, ..., m$$

$$\mu_k - \mu_{k+1} \geq d(k, \varepsilon) \quad k = 1, 2, ..., n-1$$

$$\mu_n \geq d(n, \varepsilon)$$

where $\omega_{jk}$ is the number of rank $k$ vote obtained by $DMU_j$ and $\mu_k$ is the weight of rank $k$ calculated by Model (2). Given that $\mu_k \geq \mu_{k+1}$, the extra constraint $\mu_k - \mu_{k+1} \geq d(k, \varepsilon)$ indicates the performance magnitude of vote $k+1$ versus vote $k$. The notation $d(k, \varepsilon)$ is a function which is non-decreasing in $\varepsilon$ and is referred to as a discrimination intensity function. Model (2) is solved for each candidate $j = 1, 2 ..., m$. This methodology has been widely used to evaluate alternatives in the Multi Attribute Decision Making (MADM) environment. Zerafat Angiz et al. (2012) used a modified Cook and Kress (1990) model to improve the result of the cross-ranking of DMUs in DEA. By doing so, the existing non-ordinal data were converted to the ordinal form. The following is the proposed methodology to study the influence of the inputs of DMUs in their efficiencies:

1. Consider $n$ DMUs. *To* measure the distance of inputs from the efficiency frontier, all DMUs are evaluated by the following modified DEA model derived from Model (1):

$$\Omega^*_{kp} = \max \{1 - w_{kp}\} \quad (3)$$

s.t.

$$\sum_{j=1}^{n} \lambda_j x_{ij} \leq x_{ip} \quad i = 1, 2, \dots, m; i \neq k$$

$$\sum_{j=1}^{n} \lambda_j x_{kj} \leq x_{kp} - w_{kp}$$

$$\sum_{j=1}^{n} \lambda_j y_{rj} \geq y_{rp} \quad r = 1, 2, \dots, s$$

$$\lambda_j \geq 0 \quad j = 1, 2, \dots, n$$

$$w_{kp} \geq 0$$

The optimal solution $w^*_{kp}$ indicates the distance of the $k^{th}$ input of $DMU_j$ to the efficiency frontier.

Assume that $w^*_{ij}$ is the entry of matrix $W = \left(w^*_{ij}\right)_{m \times n}$ in which $m$ and $n$ are the numbers of inputs and DMUs, respectively. Given that the number of the linear programming (LP) problems solved for each DMU, is equal to the number of inputs, there are $n \times m$ LP problems to be solved.

2. In this step, the inputs of DMUs are studied. In fact, the key purpose of this paper is to study the impact of the inputs on the efficiency of DMUs. For this, each input is considered as an alternative or candidate. Consequently, the following matrix $W'$ is introduced to convert the input data in $W$ to output data.

$$W' = \left(w'_{ij}\right)_{m \times n}, \quad w'_{ij} = 1 - \frac{w_{ij}}{\max_{j \in I}\{w_{ij}\}} \quad (4)$$

3. With reference to matrix $W' = \left(w_{ij}\right)_{m \times n}$, we construct the ranking matrix $V = \left(v_{ij}\right)_{m \times n}$ in which $v_{ij}$ is the rank of $i^{th}$ input of $DMU_j$, compared with the same inputs of other DMUs.

4. In the preferential voting framework, each candidate $i$ receives number $z_{im}$ of $m^{th}$ place votes (number $z_{i1}$ of first place votes, $z_{i2}$ of second place votes, etc.). Thus, assume that

$z_{ik}$ is the number of times that input $i^{th}$ is placed in rank $k$. Matrix $Z = (z_{ik})_{m \times m}$ is introduced with reference to matrix $V = (v_{ij})_{m \times n}$.

5. Construct matrix $B = (\beta_{ik})_{m \times m}$ where $\beta_{ik}$ is the summation of the values in matrix $W' = (w'_{ij})_{m \times n}$ which corresponds to $i^{th}$ input being placed in rank $k$. In fact, the rank place of the inputs is first obtained from matrix $V = (v_{ij})_{m \times n}$ followed by determining the corresponding value in matrix $W' = (w'_{ij})_{m \times n}$.

6. In this stage, each row in matrix $B = (\beta_{ik})_{n \times n}$ is considered as an alternative or a candidate. Given that candidate $\beta_i$ $i=1, 2, ..., m$ is representative of input $i^{th}$ of DMUs. In this research, by manipulating the Cook and Kress method (1990), we replace $z_{ik}$ by $\beta_{ik}$. The modified method, corresponding with candidate $\beta_i$ is as follows:

$$\max \ \beta_i = \sum_{k=1}^{m} \mu_k \beta_{ik} \tag{5}$$

s.t.

$$\sum_{k=1}^{m} \mu_k \beta_{ik} \leq 1 \qquad i = 1, 2, ..., m$$

$$\mu_k - \mu_{k+1} \geq d(k, \varepsilon) \quad k = 1, 2, ..., m-1$$

$$\mu_m \geq d(n, \varepsilon)$$

In the above model, the value $d(k, \varepsilon)$ is obtained using the methodology presented in Cook and Kress (1990). To clarify the philosophy behind the mathematical approach utilized in this paper, we return to a single output/single input case. Suppose there are eight DMUs which we label A to H. In Figure 2, the dotted line shows the regression line passing through the origin. Based on input-output structure governed in the data, we could define the points above the regression line as excellent and points below it as inferior or unsatisfactory. If we use DEA, we measure the degree of excellence of the best DMU ($DMU_B$) and measure the efficiency of other DMUs by its derivations from it. There thus exist a fundamental difference between statistical approaches via regression analysis and DEA. The regression form reflects "average" or "central tendency" of the observation while the latter deals with best performance and evaluates all performances by deviations from the frontier line. These two points of view can result in major differences when used as methods of evaluation.

**Figure 2:** Comparison between efficiency frontier and regression line

To study the interaction between the data, the regression approach relies on the distances of the data from the regression line. But, in an input-output structure, the model cannot distinguish between an excellent and an inferior data. For example, the effect of input *x* on output *y* is influenced by the distance of DMUs E and F to regression line, regardless of being excellent or inferior data. Interaction between the data using the methodology presented in this paper relies on the distance from the DEA efficiency frontier. The excellent data lie on the frontier to serve as a "benchmark" to use in seeking improvement.

Step 1 measures the distance of the inputs from the efficiency frontier by solving Model 3. Consider nine DMUs each with two inputs ($I_1$ and $I_2$) and single output ($O$). To illustrate production possibility set in two dimensional spaces, the output is unitized to 1 under constant-return-to-scale. Input values are normalized as indicated in Figure 3.

**Figure 3:** Measuring the distance of the inputs from the efficiency frontier in Stage 1

For instance, the distances of inputs $I_1$ and $I_2$ from the efficiency frontier corresponding to $DMU_A$ and $DMU_B$ have been shown in the above figure. The variables $w_{1A}, w_{1B}, w_{2A}$ and $w_{2B}$ illustrate the distance of first and second inputs to the efficient frontier, associated with $DMU_A$ and $DMU_B$, respectively. Step 2 deals with the closeness of the inputs to the frontier. In this study, the above-mentioned methodology is applied to rank the factors effective on growing the abundance of macrobenthos.

## 3. Sample collection and analysis

The study was performed at four locations in Penang coastal water bimonthly for a period of one year. Penang is the second smallest state of the thirteen states of Malaysia, covering approximately 1031 square kilometers. It is located on the northwestern coast of Peninsular Malaysia, bounded to the North and East by the State of Kedah, to the South by the State of Perak and to the West by the Straits of Malacca and Sumatra, Indonesia (Figure 4). The island is located in the equatorial belt between latitudes 5º 7' N and 5º 35' N and longitudes 100º 9' E and 100º 32' E (Chan, 1991).

Four sites (Teluk Bahang, Teluk Aling, Teluk Ketapang and Pantai Acheh), which are located around the North West coastal waters of Penang National Park, were selected as the sampling sites (see Figure 4). Teluk Bahang was chosen due to fisheries and ecotourism activities and Teluk Ketapang was assumed to have a minimally altered by anthropogenic activities of stress. At each site, distances of 200m, 400m, 600m, 800m, 1000m and 1200m from the shore were sampled. The characteristics of the four sites are summarized in Table 1. The coverage area of each sample was recorded by using a model 12X Garmin Global Positioning System (GPS).

**Figure 4:** Location of macrobenthic sampling stations (Penang National Park) in the coastal waters of Straits of Malacca. Transect (1=200 m, 2=400 m, 3=600 m, 4=800 m, 5=1000 m and 6=1200 m).

Sample collection of macrobenthic communities and water quality parameters were carried out bimonthly from June 2010 until April 2011. The abundance and diversity of macrobenthic community was investigated in coastal water from 200 m to a distance of 1.2 kilometer beyond the shoreline, with different depths. The sampling point abbreviations and distances from the coastline are listed in Table 1. A total of 432 samples were collected from the 4 sites where three replicates were collected at each transect. Transects 1 to 6 of each site were located approximately 200 meter away from the coastal to 1200 m towards the sea.

**Table 1:** Main attribute of the four sampling locations

Benthic samples were collected using a hand-operated 152 mm × 152 mm (6 inch × 6 inch) Ponar grab (Wildco, Cat. No 1728). The ponar grab was dropped into the seabed to grab the sediment. Each sample collection was made in triplicate. A benthos sample usually consists of a volume of sediment from which the animals were extracted. The first was carried out in the field with a view to reduce the bulk of material with a sieve of 0.5 mm (considered to be the ideal size for sampling macrofaunal organisms). The 10% of formalin and Rose Bengal were added to the samples and this preservation technique was modified from Holme and McIntyre (1971). The organisms that were found in the sample were transferred into a universal bottle which contained 70% alcohol. The benthic organisms were identified by referring to Chuang (1961), Day (1967), Chihara& Murano (1997), Fauchald (1997) and Bruyne (2003).

The measurement for water salinity, dissolved oxygen, temperature and conductivity were carried out using YSI 85 DO-SCT (Dissolved oxygen, Salinity/Conductivity/Temperature) meter. For total suspended solid (TSS), chlorophyll-*a*, phosphate, nitrite, nitrate and ammonia measurement, water samples were collected along the sampling sites using the Beta water sampler (Horizontal Transparent Acrylic 2.2L).

Particle size was determined following the method described by Eleftheriou and McIntyre (2005). The dispersed sediment suspension was washed from the beaker on to 1000μm, 425μm, 250μm, 125μm, 63μm sieve. This was carried out manually with the sieve partially immersed in washbasin containing clean water so that the sediment is submerged. Finally, the sieves and contents were transferred to an oven and were dried at 105°C. Proper weights were checked to determine the time required to achieve a constant weight. Organic matter was analyzed according to Ong *et al* (1988). Soil organic matter was determined by ashing 2.0 g of oven dried (105°C) soil in a muffle furnace at 500 °C for at least 3 hours. The loss in weight is taken as the organic matter content. The organic matter was analyzed by an ash-free dry weight method, which measured the weight loss after ignition of dry sediment at 500°C overnight (Cheng and Chang, 1999).

Over one year of bimonthly sample collection, sixty two families belonging to 26 orders and 4 phyla were identified. Mollusks were the most abundant family, comprising more than 70% of all individuals counting. The values of the physiochemical and biological factors evaluated are listed in Table 2. The study area is described by the heterogeneity of sediments. The sediments of all sampling stations contained principally mud (more than 70%) and the lowest percentage (less than 1%) was observed to be coarse sand which was ignored in this study. The near coastal regions of Teluk Bahang and Teluk Aling tend to have more mixed sand while distant locations contain more silt clay.

**Table 2:** DMUs with eleven inputs and a single output associated with each of the seven species of macrobenthos

## 4. Data analysis

In this section we would like to assess the behavior of the data and their interactivity using two methods of PCA and DEA. Figure 5 illustrates the range of values for the sampled input and output data. It is evident from the figure that among the inputs TSS, and among the outputs Gastropoda and Bivalva have the highest variation. Here, we would like to investigate the interactivity of these variables.

**Figure 5:** Distribution of the input and output variables based on sampled data

### 4.1. Principal Component Analysis (PCA)

Principal Component Analysis (PCA) using the Euclidean distance was performed to determine differences in environmental variability between transects. From the PCA result analyzed using Statistica 8 software, the first three components were measured with eigenvalues more than one (PC1 = 5.99; PC2 = 2.55 and PC3 = 1.76). Three principal components measured for 51.52% of the total variance of the original data (Table 3). According to Chatfield and Collins, (1980), influential parameters (components) are those with eigenvalues more than one in the correlation matrix which deserve to be considered in the modeling. As most data is expected by the first three components, the remaining principal components measure for a small proportion of the variance in the data and thus were negligible. Scree plot for this set of data also revealed a dramatic fall of eigenvalues from the first to the third eigenvalue and after that, the decline is gradual. As a result, the first three components could be considered significant in the analysis.

The first component (PC1) is the one that well characterized the connection between macrobenthic abundance and other few important variables as shown in Table 4 (organic matter, coarse sand, medium sand, very find sand, find sand, silt and silt-clay). These parameters (organic matter and sediment particle size) can be grouped as the sediment particle size. In the context of macrobenthic community, sediment particle size has influence on macrobenthic abundance due to their lives in the sediment. Medium sand and find sand have the component loading that was closest to macrobenthic abundance which reveals that they were closely connected (medium sand=0.387 and find sand= 38.9)

In the second component (PC2), dissolved oxygen, salinity, T.S.S and temperature are the important variable considered. The fourth important variable for PC2 is pH with fraction of 14.9%. Although important, no connection can be found between these parameters and macrobenthic abundance. The third component (PC3), where the component loading were low and furthermore, the percentage of total variance explained measured for PC3 was only 8.82%, thus no variable was significantly important. Therefore, we could group the four key parameters (dissolved oxygen, TSS, temperature, and pH) of PC2 and PC3 as the physical properties of water.

**Table 3:** Eigenvalues of Principal Components Analysis

**Table 4:** Rotated principal component loadings for 20 standardized sediment parameters and environmental factors. The three PCA factors had eigenvalues more than 1.

The fluctuation of macrobenthos at all study locations responded to the environmental parameters was calculated using Spearman correlation. The macrobenthic abundance ($r = 0.73$), mollusca ($r = 0.69$), polychaete ($r = 0.66$) and echinodermata ($r = 0.6$) significantly correlated with the first component. Other than these macrobenthic families, crustacean ($r = 0.36$) and macrobenthic diversity ($r = 0.19$) also affected of the first component significantly (Table 5). The crustacean abundance also correlated with the second component ($r = 0.27$).

**Table 5:** The Spearman Rank Correlation of macrobenthos and three principal components at all locations.

As it is evident, both of the results given in Tables 4 and 5 confirm the influence of the sediment characteristics on the macrobenthic abundance. It is however important to rank each of the input parameters based on their impact on the given output (macrobenthos species). Next, we use the DEA methodology of Section 2 and implement it for ranking of the input parameters.

### 4.2. Data Envelopment Analysis

To verify the credibility of the DEA-based algorithm proposed in Section 2, we apply DEA analysis to assess the collected data of Table 2 and compare the result with those of PCA. Consider 136 set of data points for ecological DMUs listed in Table 2. Columns 1 to 11 are environmental parameters which have impact on the macrobenthic abundance and are the inputs of this study. Columns 12 to 18 are the outputs which are the amounts of seven different macrobenthic organism including Polychaetes ($O^P$), Bivalva ($O^B$), Gastropoda ($O^G$), Scaphopoda ($O^S$), Amphipoda ($O^A$), Cumacea ($O^C$), and Echinodermata ($O^E$), respectively. In summary, for each of the benthic fauna (output), 136 DMUs with 11 inputs are available. Model (3) is therefore implemented for each of the seven macrobenthos in which $DMU_j^p$ refers for DMU number $j$ for polychaetes.

Figure 6 shows the efficiency score of environmental parameters using all gathered data of the study. According to results of DEA model between macrobenthos and sediment particle size, the most significant variable is medium sand which is important parameter for distribution of macrobenthic community. Therefore, medium sand is recorded as rank 1 followed by fine sand, very fine sand, and silt. The highest effect of medium sand is recorded for Amphipod, Cumacea and Echinodermata and the lowest influence was observed for Scaphopoda. The mean rank of very fine sand is 3 for all sampling stations. The lowest influence of fine sand is recorded for Polychaetes and Scaphopoda with rank 2. The maximum effect of silt is observed for Gastropoda with rank 6.

The range of organic matter content is within 3.03 % in Teluk Ketapang and 4.33% in Teluk Bahang. Generally the organic matter content is observed with higher rate in sampling stations near the coast. Nevertheless, samples taken near the coast reveal high macrobenthic abundance with great presence of organic matter, especially in terms of Polychaetes and Mollusca. DEA shows the maximum rank of organic matter for Echinodermata, Gastropoda and Cumacea. The average value of organic matter rank of all macrobenthos is 5. Therefore, the direct relation between organic matter and the macrobenthic abundance is evident which implies that the changes in organic matter contents influence the macrobenthic abundance across the sampling sites.

The coastal water quality of Penang National Park appears to have limited effect on the macrobenthos indicators utilized herein, although it is expected to influence macrobenthic distributions. Dissolved oxygen (DO) is found to be adequate in all sampling stations, without any anaerobic condition. This model reveals that dissolved oxygen is effective parameter among water quality with mean rank 7. The DEA model shows that the highest effect of DO is for Gastropoda and Scaphopoda. The DO values across all sites reveal small fluctuations.

According to the DEA results, the least important parameters are temperature, salinity and pH ranked 9, 10 and 11, respectively. The maximum effect of the temperature is found for Echinodermata, Gastropoda and Scaphopoda and the minimum impact is found for Amphipoda abundance. Temperature range in all sampling stations is observed between 28.83°C in Teluk Aling and 31.6°C in Teluk Ketapang. Salinity is around 28.43% in Teluk Ketapang and 30.2‰ in Teluk Aling for all the sampling stations with mean rank 10. The highest pH value is found to be 8.6 in Pantai Acheh and lowest pH is 8.16 in Teluk Bahang.

**Figure 6:** Efficiency score of environmental parameters for all outputs of study area

The macrobenthic assemblage structure and composition in the four sites are diverse due to the variations in the macrobenthos environments, sediment characteristics, scales and types of the pressures imposed on each of the four regions, as well as the various natural hydromorphological statuses. The macrobenthos quality of Teluk Aling and Teluk Bahang was better than the other locations, with higher abundance and richness.

The distinctions in the ranking of the DMUs are due to the various strategies being utilized in the techniques. Nevertheless, there appears to be some resemblance of the results. It is noteworthy that DMU of medium sand revealed the highest rank for all organisms. Medium sand, fine sand, very fine sand and silt (sediment characteristics) are the top four recommendable DMUs according to the results of this study.

**Table 6:** Results of model (5) of step 6

Tem: Temperature, DO: Dissolved Oxygen, TSS: Total Suspended Solid (TSS) , O.M: Organic matter,  M.S: Medium sand (%) 425 µm, F.S: Fine sand (%) 250 µm, VF.S: Very fine sand (%) 125 µm, S: Silt (%) 63 µm, SC: Silt and clay (%) 63≥ µm

## 4.3. Results implications

The total suspended solids revealed an increase in Pantai Acheh due to sediment particle size was mainly mud. This area is located in mangrove area with a small river flowing into the sea. Many fishing boat use this river to commute and thus causes water turbulence and high turbidity in this site. Higher TSS values may be connected with an enhancement in land perturbation activities near and the river corridor and this might be subject to bank erosion from higher peak flows and its discharge to coastal water. The macrobenthos crustacean assemblage of the surveyed regions is evidently predominated by Amphipods and Cumacea. The abundance was well related to sediment particle size, matching with the recorded by others researchers (Desroy et al., 2003; Mannino and Montagna, 1997). Based on DEA analysis, it is momentous to perceive that the highest efficiency of environmental factors was medium sand is located at rank 1. For this reason, the abundance of macrobenthos was changed from region with high particle size to area with less particle size.

Changes of water quality at these areas did not significantly influence on macrobenthos communities. According to the results of DEA, temperature, pH and salinity are ranked 10, 11 and 12, respectively. These are in line with the PCA results which found the sediment particle size as the major factor and physical properties of water as the next (less) important factor.

## 5. Conclusion

In this paper an algorithm was presented in which the performance evaluation methods were exerted to rank the inputs in a Data Envelopment Analysis model. First, DMUs were evaluated using a modified DEA model and then considering the inputs of the mentioned DMUs as alternatives (candidates), the sensitivity of the inputs was studied. To this end, a modified Cook and Kress' model was utilized to evaluate the influence of inputs in producing of the output of

DMUs. The difference between the proposed method and Cook and Kress (1990) is that the real data plays the major role in the analysis, in our method. On the other words, the real data is used rather than the number of rank place used in the Cook and Kress model. The result of the ecological study utilized in the above mathematical methodology, confirms its effectiveness and efficiency. As it was expected, the sediment characteristics obtained the higher ranks compared with the water parameters. The ranks of 1 to 5 were assigned to the sediment characteristics, and the water parameters were placed in ranked 6 to 11. The priority of the sediment characteristics over the water parameters is true for all different macrobenthic organisms. PCA methodology also revealed the sediment particle size as the major factor and physical properties of water as the next and less important factor.

**Acknowledgments**

The authors would like to thank the officers and crew of Centre of Marine and Coastal Studies (CEMCS), Universiti Sains Malaysia research vessel for help in on board sample collection. This project was funded by the University Research Grant, Universiti Sains Malaysia (Account number: 1001/PPANTAS/815052).

**References**

Adler, N., Friedman, L., & Sinuany-Stern, Z. (2002). Review of ranking methods in the data envelopment analysis context. *European Journal of Operational Research*, *140*(2), 249–265.

Bakalem, A., Ruellet, T., & Dauvin, J. C. (2009). Benthic indices and ecological quality of shallow Algeria fine sand community. *Ecological Indicators*, *9*(3), 395–408.

Chan, N.W. (1991). The climate of Penang Island. Kajian Malaysia, Jil. IX, 1: 62-86.

CHAPIN, F. S., CHAPIN, C., MATSON, P. A. & VITOUSEK, P. 2011. *Principles of Terrestrial Ecosystem Ecology*, Springer New York.

Clark, R.B.(1997). Marine Pollution (4th Edn). Oxford: Clarendon Press.

Cook, W. D., & Kress, M. (1990). A data envelopment model for aggregating preference rankings. *Management Science*, *36*(11), 1302–1310.

Costa, H. H., G. A. Weerasundara and A. Pathiratne. 1997. Species composition, abundance and distribution of aquatic oligochaetes in Colombo (Beira) lake, Sri Lanka. Sri Lanka Journal of Aquatic Science 2:69-80.

Cowell, B. C., Dye, C. W., & Adams, R. C. (1975). A synoptic study of the limnology of Lake Thonotosassa, Florida. *Hydrobiologia*, *46*(2-3), 301–345.

CUFF, D. & GOUDIE, A. 2009. *The Oxford Companion to Global Change*, Oxford University Press, USA.

HILLEWAERT, H. 2006. Scheme of eutrophication. *In:* EUTROPHICATION.JPG (ed.).

Dauer, D. M., Ranasinghe, J. A., & Weisberg, S. B. (2000). Relationships between benthic community condition, water quality, sediment quality, nutrient loads, and land use patterns in Chesapeake Bay. *Estuaries*, *23*(1), 80–96.

Desroy, N., Warembourg, C., Dewarumez, J. M., & Dauvin, J. C. (2003). Macrobenthic resources of the shallow soft-bottom sediments in the eastern English Channel and southern North Sea. *ICES Journal of Marine Science: Journal du Conseil*, *60*(1), 120–131.

Eleftheriou, A., McIntyre, A.D. (2005). Methods for the study of marine benthos. Blackwell Science.

Gamlath, G.A.R.K. and M.J.S. Wijeyaratne.1997. Indicator organisms of environmental conditions in a lotic water body in Sri Lanka - Short communication. *Sri Lanka Journal of Aquatic Sciences*, 2: 121 - 129.

Hashimoto, A. (1997). A ranked voting system using a DEA/AR exclusion model: A note. *European Journal of Operational Research*, *97*(3), 600–604.

Hellawell, J. M. (1986). Biological indicators of freshwater pollution and environmental management.

HOSSEINZADEH LOTFI, F., JAHANSHAHLOO, G. R., KHODABAKHSHI, M., ROSTAMY-MALKHLIFEH, M., MOGHADDAS, Z. & VAEZ-GHASEMI, M. 2013. A Review of Ranking Models in Data Envelopment Analysis. *Journal of Applied Mathematics,* 2013**,** 20.

IPCC 2014. *Climate Change 2014: Impacts, Adaptation, and Vulnerability. Part A: Global and Sectoral Aspects. Contribution of Working Group II to the Fifth Assessment Report of the Intergovernmental Panel on Climate Change [Field, C.B., V.R. Barros, D.J. Dokken, K.J. Mach, M.D. Mastrandrea, T.E. Bilir, M. Chatterjee, K.L. Ebi, Y.O. Estrada, R.C. Genova, B. Girma, E.S. Kissel, A.N. Levy, S. MacCracken, P.R. Mastrandrea, and L.L. White (eds.)],* Cambridge, United Kingdom and New York, NY, USA, Cambridge University Press.

Jackson, J. B. C., Kirby, M. X., Berger, W. H., Bjorndal, K. A., Botsford, L. W., Bourque, B. J., … Estes, J. A. (2001). Historical overfishing and the recent collapse of coastal ecosystems. *science*, *293*(5530), 629–637.

Jan, R.Q., Dai, C.F., and Chang, K.H., 1994. "Monitoring of Hard Substrate Communities," Biomonitoring of Coastal Waters and Estuaries, Kramer, K.J.M. (ed.), CRC Press, Boca Raton, FL, pp. 285-310.

Kennish, M.J. (1997): Practical Handbook of Estuarine and Marine Pollution. CRC Press, Boca Raton, FL.

Leopold, A. (1933). The conservation ethic. *Journal of Forestry*, *31*(6), 634–643.

Mannino, A., & Montagna, P. A. (1997). Small-scale spatial variation of macrobenthic community structure. *Estuaries*, *20*(1), 159–173.

Muxika, I., Borja, A., & Bald, J. (2007). Using historical data, expert judgement and multivariate analysis in assessing reference conditions and benthic ecological status, according to the European Water Framework Directive. *Marine Pollution Bulletin*, *55*(1), 16–29.

Ong, B., Omar, A. and Wan Din, W.M. (1988) Hidrobiologi Perairan Pulau Pinang. Universiti Sains Malaysia.1-39.

Pathiratne, A., Weerasundara, A., 2004. Bioassessment of selected inland water bodies in Sri Lanka using benthic oligochaetes with consideration of temporal variations. *International review of hydrobiology*. 89, 305-316.

Saati, S., M. Zarafat Angiz, A. Memariani and G.R. Jahanshahloo. (2001). A model for ranking decision making units in data envelopment analysis. *Ricerca Operativa*. 31(97): 47- 59.

Warwick, R., 2006. Environmental impact studies on marine communities: pragmatical considerations. *Australian Journal of Ecology*. 18, 63-80.

Weerasundara, A., Pathiratne, A., & Costa, H. (2000). Species composition and abundance of littoral oligochaete fauna in Lunuwila reservoir, Sri Lanka. *International review of hydrobiology*, *85*, 223–230.

Zerafat Angiz, M., Mustafa, A., Kamali, M.J., 2012. Cross-ranking of Decision Making Units in Data Envelopment Analysis. *Applied Mathematical Modelling*.

Zerafat Angiz L, M., Mustafa, A., Abdul Ghani, N., Kamil, A. A., 2012. Group Decision via Usage of Analytic Hierarchy Process and Preference Aggregation Method. Sains Malaysiana 41, 361-366.

**Table A1.** DMUs with a single output associated with each species of macrobenthos and eleven inputs

| | Polychaeta ($O^P$) | Bivalva ($O^B$) | Gastropoda, ($O^G$) | Scaphopod a ($O^S$) | Amphipod ($O^A$) | Cumacea ($O^C$) | Echinodermat a ($O^E$) | Tem ($I_1$) | DO ($I_2$) | pH ($I_3$) | Salinity ($I_4$) | TSS ($I_5$) | O.M ($I_6$) | M.S ($I_7$) | F.S ($I_8$) | VF.S ($I_9$) | S ($I_{10}$) | SC ($I_{11}$) |
|---|---|---|---|---|---|---|---|---|---|---|---|---|---|---|---|---|---|---|
| 1 | 2 | 48 | 12 | 0 | 1 | 3 | 1 | 31.2 | 5.15 | 8.2 | 30 | 100.4 | 3 | 6.40 | 3.74 | 2.72 | 2.29 | 81.58 |
| 2 | 2 | 3 | 3 | 4 | 1 | 3 | 0 | 31 | 5.6 | 8.2 | 30 | 120 | 3.5 | 0.33 | 0.47 | 0.93 | 1.50 | 96.77 |
| 3 | 1 | 4 | 18 | 0 | 1 | 5 | 1 | 31.5 | 5.58 | 8.3 | 30 | 77.2 | 2.9 | 0.18 | 0.15 | 0.78 | 1.47 | 97.42 |
| 4 | 6 | 19 | 25 | 0 | 4 | 1 | 0 | 31.2 | 5.75 | 8.3 | 30.11 | 152 | 2.25 | 0.06 | 0.23 | 0.82 | 1.80 | 97.09 |
| 5 | 2 | 4 | 4 | 0 | 1 | 2 | 1 | 31.2 | 4.93 | 8.3 | 30 | 74.4 | 3 | 0.09 | 0.13 | 0.89 | 1.48 | 97.42 |
| 6 | 45 | 8 | 27 | 0 | 31 | 19 | 2 | 31.1 | 4.5 | 8.3 | 30 | 85.2 | 2.5 | 3.26 | 6.00 | 8.07 | 8.17 | 73.83 |
| 7 | 1 | 0 | 11 | 0 | 4 | 18 | 3 | 31.1 | 4.46 | 8.3 | 30.2 | 95.6 | 1.25 | 0.17 | 0.44 | 1.21 | 1.36 | 96.82 |
| 8 | 4 | 15 | 12 | 2 | 10 | 24 | 0 | 31.2 | 4.7 | 8.3 | 29.9 | 88.4 | 2 | 0.14 | 0.29 | 0.59 | 1.35 | 97.63 |
| 9 | 5 | 5 | 12 | 0 | 4 | 5 | 0 | 31.3 | 4.93 | 8.3 | 29.9 | 63.2 | 2.5 | 0.04 | 0.21 | 0.79 | 0.83 | 98.12 |
| 10 | 3 | 29 | 14 | 0 | 2 | 1 | 1 | 31.4 | 5.62 | 8.3 | 29.9 | 56.8 | 1.6 | 0.07 | 0.37 | 1.53 | 2.22 | 95.81 |
| 11 | 1 | 54 | 25 | 0 | 2 | 4 | 0 | 31.4 | 4.98 | 8.4 | 30 | 68 | 2.5 | 0.05 | 0.08 | 0.40 | 0.85 | 98.62 |
| 12 | 10 | 7 | 39 | 0 | 5 | 5 | 2 | 31.1 | 5.24 | 8.2 | 29.9 | 105.2 | 3.5 | 0.07 | 0.09 | 0.30 | 0.80 | 98.74 |
| 13 | 2 | 3 | 10 | 0 | 5 | 7 | 0 | 31.2 | 5.29 | 8.3 | 30 | 81.4 | 1.3 | 0.06 | 0.10 | 0.22 | 0.41 | 99.21 |
| 14 | 1 | 1 | 3 | 0 | 5 | 4 | 0 | 31.3 | 5.22 | 8.3 | 30 | 220 | 2.5 | 0.04 | 0.20 | 0.50 | 2.17 | 97.09 |
| 15 | 2 | 5 | 22 | 0 | 8 | 14 | 1 | 31.3 | 5 | 8.3 | 30 | 56.6 | 1.5 | 0.11 | 0.04 | 0.36 | 0.76 | 98.73 |
| 16 | 5 | 18 | 34 | 0 | 2 | 4 | 2 | 31.3 | 5.4 | 8.3 | 30 | 68.6 | 3.03 | 0.08 | 0.44 | 1.42 | 0.87 | 97.19 |
| 17 | 1 | 10 | 10 | 0 | 1 | 1 | 0 | 31.3 | 5.1 | 8.3 | 30 | 87.6 | 2 | 0.18 | 0.21 | 0.33 | 0.90 | 98.37 |
| 18 | 6 | 71 | 16 | 5 | 5 | 6 | 1 | 31.6 | 5.88 | 8.3 | 29.4 | 95.6 | 4.33 | 0.10 | 0.13 | 0.65 | 0.97 | 98.15 |
| 19 | 9 | 9 | 6 | 0 | 5 | 6 | 0 | 31.7 | 6.03 | 8.4 | 29.5 | 79.6 | 4 | 0.04 | 0.09 | 0.41 | 0.66 | 98.80 |
| 20 | 1 | 7 | 18 | 1 | 2 | 1 | 1 | 31.6 | 5.18 | 8.4 | 29.8 | 92.4 | 3.8 | 0.04 | 0.09 | 0.32 | 1.01 | 98.53 |
| 21 | 1 | 11 | 11 | 0 | 1 | 1 | 0 | 31.6 | 4.8 | 8.3 | 29.9 | 62.8 | 2 | 0.09 | 0.14 | 0.40 | 1.04 | 98.33 |
| 22 | 11 | 8 | 1 | 0 | 1 | 3 | 1 | 31.6 | 4.85 | 8.3 | 30 | 74.4 | 2.2 | 0.04 | 0.06 | 0.74 | 1.18 | 97.98 |
| 23 | 4 | 19 | 17 | 2 | 1 | 4 | 0 | 31.6 | 4.62 | 8.3 | 30 | 98 | 2 | 0.05 | 0.07 | 0.44 | 1.08 | 98.36 |
| 24 | 1 | 130 | 11 | 7 | 2 | 4 | 1 | 30 | 6.8 | 8.4 | 30.1 | 79.4 | 2.5 | 9.41 | 5.50 | 4.01 | 3.30 | 73.02 |
| 25 | 2 | 1 | 16 | 0 | 1 | 2 | 0 | 30 | 6.49 | 8.4 | 30.15 | 87 | 2.1 | 0.60 | 2.04 | 1.57 | 1.13 | 94.67 |
| 26 | 2 | 7 | 14 | 2 | 1 | 1 | 1 | 30.1 | 5.5 | 8.4 | 30.07 | 85.2 | 2 | 0.32 | 0.43 | 0.46 | 2.43 | 96.36 |
| 27 | 2 | 12 | 36 | 0 | 1 | 1 | 0 | 30 | 6.42 | 8.4 | 30.1 | 89.2 | 1.4 | 0.18 | 0.11 | 0.72 | 1.04 | 97.94 |
| 28 | 1 | 3 | 9 | 2 | 1 | 1 | 0 | 29.9 | 6.01 | 8.4 | 30.1 | 112.4 | 1.6 | 0.61 | 2.01 | 1.41 | 1.03 | 94.94 |
| 29 | 4 | 6 | 16 | 1 | 3 | 6 | 1 | 29.9 | 6.08 | 8.4 | 30.1 | 92.4 | 1.9 | 2.02 | 0.97 | 2.06 | 1.21 | 93.74 |
| 30 | 3 | 4 | 13 | 0 | 2 | 3 | 0 | 29.8 | 5.8 | 8.4 | 29.9 | 84.8 | 1.3 | 0.44 | 1.13 | 1.65 | 1.08 | 95.71 |
| 31 | 0 | 8 | 19 | 2 | 0 | 1 | 0 | 29.7 | 5.64 | 8.3 | 30 | 82 | 1 | 0.25 | 0.44 | 0.79 | 0.93 | 97.59 |
| 32 | 1 | 134 | 15 | 1 | 1 | 1 | 0 | 29.8 | 5.35 | 8.4 | 30 | 65 | 1.25 | 0.58 | 2.80 | 4.61 | 2.25 | 89.75 |
| 33 | 1 | 15 | 19 | 4 | 1 | 1 | 1 | 29.8 | 5.9 | 8.3 | 30 | 94.8 | 2 | 0.32 | 0.28 | 1.08 | 0.93 | 97.40 |
| 34 | 3 | 1 | 2 | 0 | 2 | 2 | 1 | 29.8 | 6.4 | 8.4 | 30 | 81.2 | 2.5 | 0.07 | 0.09 | 0.30 | 0.80 | 98.74 |
| 35 | 34 | 2 | 10 | 0 | 19 | 50 | 0 | 29.9 | 6.23 | 8.5 | 29.3 | 104 | 1.6 | 0.04 | 0.09 | 0.41 | 0.66 | 98.81 |
| 36 | 1 | 5 | 9 | 0 | 3 | 2 | 1 | 29.4 | 6.11 | 8.4 | 29.4 | 73.2 | 2.5 | 0.04 | 0.20 | 0.50 | 2.17 | 97.09 |
| 37 | 4 | 5 | 8 | 0 | 0 | 0 | 0 | 30 | 6.4 | 8.4 | 29.8 | 99.8 | 3.5 | 0.11 | 0.04 | 0.36 | 0.76 | 98.73 |
| 38 | 1 | 22 | 8 | 2 | 1 | 0 | 0 | 29.9 | 5.61 | 8.4 | 29.8 | 87 | 1.3 | 0.08 | 0.44 | 1.42 | 0.87 | 97.19 |
| 39 | 1 | 16 | 41 | 2 | 1 | 4 | 0 | 29.8 | 5.68 | 8.3 | 29.7 | 103.2 | 2.5 | 0.16 | 0.13 | 0.33 | 0.90 | 98.47 |
| 40 | 19 | 4 | 4 | 0 | 22 | 6 | 1 | 29.3 | 4.45 | 8.16 | 29.4 | 101 | 1.5 | 0.10 | 0.13 | 0.66 | 0.97 | 98.14 |
| 41 | 7 | 4 | 49 | 0 | 0 | 7 | 1 | 29.4 | 5.59 | 8 | 29.3 | 52 | 3 | 0.04 | 0.09 | 0.41 | 0.66 | 98.81 |
| 42 | 2 | 9 | 7 | 0 | 2 | 9 | 0 | 29.6 | 5.65 | 8.3 | 29.5 | 102.4 | 2 | 0.03 | 0.09 | 0.24 | 1.01 | 98.62 |
| 43 | 1 | 3 | 2 | 0 | 1 | 0 | 0 | 29.7 | 5.83 | 8.3 | 29.7 | 59 | 4.1 | 0.05 | 0.14 | 0.39 | 1.04 | 98.37 |
| 44 | 0 | 13 | 13 | 0 | 0 | 2 | 0 | 29.7 | 5.76 | 8.4 | 29.6 | 71 | 4 | 0.04 | 0.06 | 0.75 | 1.18 | 97.98 |
| 45 | 1 | 8 | 17 | 0 | 5 | 1 | 1 | 29.8 | 5.66 | 8.3 | 29.6 | 104 | 3.8 | 0.04 | 0.07 | 0.48 | 1.08 | 98.33 |
| 46 | 6 | 18 | 18 | 2 | 3 | 7 | 2 | 30.4 | 5.29 | 8.1 | 29.7 | 66.8 | 2 | 0.72 | 1.11 | 3.20 | 2.15 | 92.66 |
| 47 | 1 | 10 | 154 | 0 | 2 | 2 | 3 | 30.4 | 5.85 | 8.3 | 29.7 | 64 | 2.2 | 0.20 | 0.37 | 1.10 | 2.57 | 95.77 |
| 48 | 3 | 3 | 4 | 1 | 1 | 1 | 0 | 30.4 | 5.44 | 8.3 | 29.8 | 55.6 | 2 | 0.17 | 0.13 | 0.77 | 1.47 | 97.47 |
| 49 | 1 | 16 | 29 | 0 | 0 | 1 | 1 | 30.3 | 5.35 | 8.3 | 29.8 | 84.4 | 2.5 | 0.04 | 0.20 | 0.80 | 1.79 | 97.16 |
| 50 | 2 | 4 | 42 | 1 | 1 | 1 | 1 | 30.3 | 5.5 | 8.4 | 29.7 | 61.8 | 2.1 | 0.08 | 0.13 | 0.89 | 1.48 | 97.42 |
| 51 | 5 | 11 | 53 | 4 | 2 | 1 | 2 | 30.3 | 5.82 | 8.3 | 29.6 | 83 | 2 | 3.26 | 6.00 | 8.07 | 8.17 | 73.83 |
| 52 | 3 | 11 | 39 | 1 | 2 | 1 | 1 | 30.4 | 5.65 | 8.4 | 29.6 | 85.2 | 3 | 0.17 | 0.36 | 1.18 | 1.28 | 97.01 |
| 53 | 7 | 7 | 18 | 0 | 1 | 4 | 0 | 30.3 | 5.5 | 8.3 | 29.7 | 74.2 | 3.5 | 0.11 | 0.25 | 0.57 | 1.31 | 97.75 |
| 54 | 0 | 6 | 11 | 0 | 0 | 2 | 0 | 30.4 | 4.84 | 8.2 | 29.6 | 75 | 2.9 | 0.03 | 0.21 | 0.76 | 0.78 | 98.22 |
| 55 | 4 | 14 | 17 | 1 | 1 | 0 | 0 | 30.3 | 4.91 | 8.3 | 29.6 | 71 | 2.25 | 0.07 | 0.37 | 1.53 | 2.22 | 95.81 |
| 56 | 1 | 10 | 14 | 0 | 0 | 1 | 1 | 30.2 | 4.81 | 8.3 | 29.4 | 95.2 | 3 | 0.05 | 0.08 | 0.40 | 0.85 | 98.62 |
| 57 | 2 | 14 | 9 | 0 | 1 | 1 | 2 | 30.9 | 5.24 | 8.5 | 29.3 | 80 | 2.5 | 0.09 | 0.19 | 1.83 | 6.47 | 91.42 |
| 58 | 1 | 5 | 9 | 1 | 0 | 1 | 0 | 30.9 | 5.78 | 8.5 | 29.5 | 63 | 1.25 | 0.04 | 0.39 | 1.33 | 6.31 | 91.93 |
| 59 | 1 | 8 | 6 | 1 | 1 | 0 | 1 | 30.8 | 5.8 | 8.4 | 29.5 | 80 | 2 | 0.06 | 0.17 | 0.82 | 3.67 | 95.28 |
| 60 | 0 | 16 | 13 | 0 | 0 | 0 | 0 | 30.8 | 5.99 | 8.5 | 29.6 | 60 | 2.5 | 0.05 | 0.47 | 1.97 | 4.82 | 92.70 |
| 61 | 0 | 5 | 5 | 1 | 1 | 1 | 0 | 30.8 | 5.86 | 8.5 | 29.7 | 68 | 1.6 | 0.13 | 0.99 | 1.28 | 5.11 | 92.49 |
| 62 | 1 | 36 | 4 | 0 | 1 | 1 | 1 | 30.8 | 6.2 | 8.5 | 29.7 | 80 | 2.5 | 0.42 | 0.49 | 0.72 | 5.23 | 93.14 |
| 63 | 37 | 11 | 6 | 0 | 2 | 4 | 1 | 31 | 6.88 | 7.7 | 30 | 98 | 3.5 | 0.04 | 0.14 | 0.91 | 2.54 | 96.38 |
| 64 | 3 | 5 | 3 | 0 | 1 | 2 | 0 | 30.9 | 7.45 | 8.5 | 29.9 | 101 | 1.3 | 0.04 | 0.08 | 0.42 | 0.66 | 98.80 |
| 65 | 1 | 69 | 33 | 0 | 1 | 1 | 0 | 30.8 | 5.99 | 8.6 | 29.6 | 66 | 2.5 | 0.09 | 0.23 | 0.59 | 8.06 | 91.02 |
| 66 | 0 | 16 | 6 | 0 | 0 | 1 | 0 | 30.9 | 7.42 | 8.6 | 29.7 | 76 | 1.5 | 0.18 | 0.11 | 0.64 | 5.05 | 94.02 |
| 67 | 1 | 14 | 3 | 1 | 1 | 2 | 1 | 30.8 | 6.99 | 8.5 | 29.5 | 220 | 3 | 0.04 | 0.07 | 0.76 | 1.16 | 97.98 |
| 68 | 2 | 40 | 25 | 3 | 1 | 1 | 1 | 31.4 | 7.01 | 8.5 | 29.3 | 200 | 2 | 0.07 | 0.55 | 2.81 | 4.54 | 92.04 |
| 69 | 3 | 3 | 94 | 2 | 1 | 1 | 1 | 29.1 | 6.3 | 8.6 | 29.3 | 49.8 | 4.1 | 0.72 | 1.11 | 3.20 | 2.15 | 92.66 |
| 70 | 3 | 7 | 113 | 1 | 3 | 2 | 1 | 29.1 | 6.2 | 8.6 | 29.3 | 62.6 | 4 | 0.20 | 0.37 | 1.10 | 2.57 | 95.77 |
| 71 | 0 | 3 | 27 | 0 | 1 | 1 | 0 | 28.9 | 5.8 | 8.6 | 29.4 | 77.4 | 3.8 | 0.17 | 0.13 | 0.77 | 1.47 | 97.47 |
| 72 | 2 | 7 | 28 | 1 | 2 | 4 | 0 | 28.89 | 5.6 | 8.6 | 29.5 | 64.8 | 2 | 0.04 | 0.21 | 0.81 | 1.79 | 97.16 |
| 73 | 3 | 12 | 14 | 0 | 1 | 1 | 1 | 28.92 | 5.4 | 8.5 | 29.3 | 60.6 | 2.2 | 0.08 | 0.13 | 0.89 | 1.48 | 97.42 |

| | | | | | | | | | | | | | | | | | |
|---|---|---|---|---|---|---|---|---|---|---|---|---|---|---|---|---|---|
| 74 | 6 | 45 | 29 | 2 | 3 | 5 | 2 | 29.1 | 6.1 | 8.5 | 29.2 | 86.6 | 2 | 3.26 | 6.00 | 8.07 | 10.17 | 73.83 |
| 75 | 5 | 20 | 31 | 2 | 2 | 13 | 1 | 29.2 | 6.2 | 8.4 | 29.1 | 63.6 | 2.5 | 0.17 | 0.36 | 1.18 | 1.60 | 97.01 |
| 76 | 2 | 7 | 18 | 1 | 2 | 4 | 1 | 28.9 | 5.4 | 8.6 | 29 | 73.8 | 2.1 | 0.11 | 0.25 | 0.57 | 1.63 | 97.75 |
| 77 | 2 | 6 | 5 | 1 | 1 | 4 | 0 | 28.83 | 5.5 | 8.6 | 29.2 | 66.2 | 2 | 0.04 | 0.20 | 0.76 | 0.98 | 98.22 |
| 78 | 2 | 16 | 22 | 1 | 5 | 1 | 1 | 29.1 | 5.4 | 8.6 | 29.2 | 63 | 1.4 | 0.07 | 0.37 | 1.53 | 2.77 | 95.81 |
| 79 | 3 | 12 | 21 | 2 | 8 | 2 | 0 | 29.3 | 6.1 | 8.6 | 29.2 | 56 | 1.6 | 0.05 | 0.08 | 0.40 | 1.07 | 98.62 |
| 80 | 3 | 3 | 107 | 2 | 2 | 5 | 1 | 30.9 | 8.24 | 8.5 | 29.3 | 170 | 1.9 | 0.09 | 0.19 | 1.83 | 6.47 | 91.42 |
| 81 | 1 | 5 | 23 | 0 | 8 | 9 | 1 | 30.9 | 8.78 | 8.5 | 29.5 | 121 | 1.3 | 0.04 | 0.39 | 1.33 | 6.31 | 91.93 |
| 82 | 6 | 11 | 9 | 0 | 3 | 1 | 0 | 30.8 | 8.8 | 8.4 | 29.5 | 112 | 1 | 0.06 | 0.17 | 0.82 | 3.67 | 95.28 |
| 83 | 5 | 16 | 25 | 0 | 5 | 9 | 0 | 30.8 | 8.99 | 8.5 | 29.6 | 135 | 1.25 | 0.05 | 0.47 | 1.97 | 4.54 | 92.70 |
| 84 | 1 | 17 | 131 | 0 | 1 | 5 | 1 | 30.8 | 8.86 | 8.5 | 29.7 | 145 | 2 | 0.13 | 0.99 | 1.28 | 5.11 | 92.49 |
| 85 | 3 | 28 | 145 | 0 | 9 | 6 | 1 | 30.8 | 9.2 | 8.5 | 29.7 | 130 | 2.5 | 0.42 | 0.49 | 0.72 | 5.23 | 93.14 |
| 86 | 13 | 6 | 4 | 0 | 2 | 2 | 2 | 31 | 9.88 | 7.7 | 30 | 64 | 1.6 | 0.08 | 0.14 | 0.87 | 2.54 | 96.38 |
| 87 | 4 | 5 | 9 | 0 | 3 | 1 | 1 | 30.9 | 10.45 | 8.5 | 29.9 | 85 | 2.5 | 0.04 | 0.08 | 0.42 | 0.66 | 98.80 |
| 88 | 3 | 10 | 16 | 0 | 3 | 2 | 0 | 30.8 | 8.99 | 8.6 | 29.6 | 90 | 3.5 | 0.10 | 0.23 | 0.59 | 8.06 | 91.02 |
| 89 | 0 | 5 | 11 | 0 | 0 | 1 | 0 | 30.9 | 10.42 | 8.6 | 29.7 | 93 | 1.3 | 0.18 | 0.12 | 0.64 | 5.05 | 94.02 |
| 90 | 2 | 12 | 19 | 1 | 1 | 1 | 1 | 30.8 | 9.99 | 8.5 | 29.5 | 75 | 2.5 | 0.04 | 0.06 | 0.76 | 1.16 | 97.98 |
| 91 | 2 | 19 | 25 | 0 | 1 | 1 | 1 | 31.4 | 10.01 | 8.5 | 29.3 | 60 | 1.5 | 0.07 | 0.55 | 2.81 | 4.54 | 92.04 |
| 92 | 7 | 29 | 65 | 4 | 4 | 13 | 2 | 29.6 | 5.6 | 8.5 | 29.3 | 76 | 3 | 9.41 | 5.44 | 4.01 | 3.30 | 73.02 |
| 93 | 2 | 9 | 72 | 0 | 3 | 8 | 1 | 29.6 | 5.8 | 8.5 | 29.4 | 98.4 | 2 | 0.60 | 2.04 | 1.57 | 1.13 | 94.67 |
| 94 | 5 | 10 | 60 | 0 | 1 | 6 | 0 | 29.5 | 5.8 | 8.6 | 29.1 | 104.8 | 4.1 | 0.39 | 0.36 | 0.45 | 2.42 | 96.36 |
| 95 | 4 | 9 | 51 | 1 | 2 | 7 | 0 | 29.6 | 5.6 | 8.6 | 29 | 91.6 | 4 | 0.18 | 0.11 | 0.73 | 1.03 | 97.94 |
| 96 | 2 | 12 | 52 | 0 | 1 | 6 | 1 | 29.5 | 5.6 | 8.6 | 29.1 | 97.2 | 3.8 | 0.62 | 2.00 | 1.42 | 1.02 | 94.94 |
| 97 | 6 | 55 | 50 | 3 | 5 | 18 | 2 | 29.5 | 5.4 | 8.6 | 29.3 | 70.8 | 2 | 2.02 | 0.97 | 2.06 | 1.21 | 93.74 |
| 98 | 5 | 17 | 82 | 0 | 4 | 18 | 1 | 29.5 | 5.4 | 8.6 | 29.4 | 97.2 | 2.2 | 0.60 | 1.13 | 1.65 | 1.08 | 95.71 |
| 99 | 6 | 9 | 19 | 0 | 1 | 8 | 0 | 29.5 | 5.6 | 8.6 | 29.3 | 66 | 2 | 0.39 | 0.44 | 0.79 | 0.93 | 97.59 |
| 100 | 1 | 5 | 15 | 0 | 1 | 7 | 1 | 29.5 | 5.4 | 8.6 | 29.3 | 78.4 | 2.5 | 0.18 | 2.80 | 4.61 | 2.25 | 89.75 |
| 101 | 1 | 12 | 44 | 0 | 1 | 1 | 1 | 29.5 | 6 | 8.6 | 29.3 | 104 | 2.1 | 0.62 | 0.28 | 1.08 | 0.93 | 97.40 |
| 102 | 3 | 18 | 12 | 1 | 1 | 3 | 2 | 29.7 | 5.4 | 8.5 | 29.1 | 70.2 | 2 | 0.07 | 0.09 | 0.30 | 0.80 | 98.74 |
| 103 | 6 | 26 | 13 | 0 | 6 | 4 | 1 | 29.8 | 5.5 | 8.5 | 29.1 | 152.8 | 1.25 | 0.05 | 0.10 | 0.22 | 0.41 | 99.21 |
| 104 | 1 | 12 | 8 | 0 | 3 | 1 | 1 | 30 | 5.5 | 8.6 | 29 | 139.6 | 2 | 0.04 | 0.20 | 0.50 | 2.17 | 97.09 |
| 105 | 2 | 6 | 8 | 0 | 2 | 1 | 0 | 30 | 5.7 | 8.6 | 29.1 | 124.8 | 2.5 | 0.11 | 0.04 | 0.36 | 0.76 | 98.73 |
| 106 | 2 | 29 | 24 | 1 | 2 | 4 | 0 | 29.9 | 5.8 | 8.6 | 29.2 | 130 | 1.6 | 0.08 | 0.44 | 1.42 | 0.87 | 97.19 |
| 107 | 4 | 14 | 30 | 0 | 5 | 5 | 1 | 30 | 5.7 | 8.6 | 29.2 | 94.4 | 2.5 | 0.12 | 0.17 | 0.33 | 0.90 | 98.47 |
| 108 | 6 | 6 | 14 | 0 | 1 | 4 | 2 | 29.8 | 5.9 | 8.5 | 28.9 | 83.2 | 3.5 | 0.10 | 0.13 | 0.66 | 0.97 | 98.14 |
| 109 | 3 | 31 | 37 | 2 | 1 | 2 | 1 | 29.9 | 5.8 | 8.5 | 28.9 | 108 | 1.3 | 0.04 | 0.09 | 0.40 | 0.65 | 98.81 |
| 110 | 2 | 29 | 19 | 0 | 2 | 2 | 0 | 30 | 6 | 8.5 | 28.9 | 119.2 | 2.5 | 0.03 | 0.09 | 0.24 | 1.01 | 98.62 |
| 111 | 1 | 32 | 15 | 0 | 0 | 1 | 0 | 30 | 5.8 | 8.4 | 28.9 | 102.8 | 1.5 | 0.06 | 0.14 | 0.39 | 1.04 | 98.37 |
| 112 | 1 | 97 | 53 | 1 | 1 | 1 | 1 | 29.8 | 5.7 | 8.5 | 29 | 131.2 | 3 | 0.04 | 0.06 | 0.74 | 1.18 | 97.98 |
| 113 | 1 | 157 | 83 | 0 | 1 | 1 | 1 | 29.8 | 5.6 | 8.5 | 29.1 | 125.2 | 2 | 0.04 | 0.07 | 0.48 | 1.08 | 98.33 |
| 114 | 17 | 10 | 48 | 9 | 3 | 8 | 2 | 30.8 | 5.7 | 8.3 | 29.1 | 120.4 | 4.1 | 6.80 | 3.74 | 2.72 | 2.29 | 81.58 |
| 115 | 3 | 20 | 43 | 5 | 1 | 2 | 1 | 30.7 | 6 | 8.4 | 29.2 | 54.8 | 4 | 0.33 | 0.47 | 0.93 | 1.50 | 96.77 |
| 116 | 1 | 12 | 39 | 2 | 1 | 1 | 1 | 30.8 | 5.8 | 8.4 | 29 | 112.4 | 3.8 | 0.18 | 0.15 | 0.78 | 1.47 | 97.42 |
| 117 | 3 | 11 | 44 | 1 | 1 | 3 | 1 | 30.8 | 5.7 | 8.4 | 29 | 92.8 | 2 | 0.06 | 0.23 | 0.82 | 1.80 | 97.09 |
| 118 | 4 | 22 | 49 | 3 | 1 | 4 | 1 | 30.8 | 5.5 | 8.3 | 29.1 | 111.6 | 2.2 | 0.09 | 0.13 | 0.89 | 1.48 | 97.42 |
| 119 | 18 | 22 | 33 | 5 | 53 | 18 | 3 | 30.9 | 6 | 8.4 | 29 | 78.8 | 2 | 3.26 | 6.00 | 8.07 | 8.17 | 73.83 |
| 120 | 5 | 14 | 13 | 5 | 10 | 14 | 2 | 30.8 | 6.1 | 8.4 | 29.1 | 72.4 | 2.5 | 0.17 | 0.44 | 1.21 | 1.36 | 96.82 |
| 121 | 3 | 6 | 20 | 0 | 1 | 4 | 1 | 30.7 | 6.4 | 8.4 | 29.1 | 77.2 | 2.1 | 0.14 | 0.29 | 0.59 | 1.35 | 97.63 |
| 122 | 1 | 5 | 9 | 0 | 2 | 1 | 1 | 30.8 | 5.8 | 8.3 | 29 | 61.2 | 2 | 0.04 | 0.21 | 0.79 | 0.83 | 98.12 |
| 123 | 3 | 29 | 52 | 1 | 1 | 8 | 1 | 30.8 | 5.2 | 8.4 | 29.1 | 93.6 | 3.5 | 0.07 | 0.37 | 1.53 | 2.22 | 95.81 |
| 124 | 8 | 33 | 59 | 0 | 1 | 14 | 2 | 30.8 | 5.8 | 8.3 | 29 | 134.8 | 1.3 | 0.05 | 0.08 | 0.40 | 0.84 | 98.62 |
| 125 | 1 | 11 | 14 | 0 | 1 | 1 | 1 | 30.9 | 6.3 | 8.4 | 28.8 | 95.6 | 2.5 | 0.07 | 0.09 | 0.30 | 0.80 | 98.74 |
| 126 | 3 | 8 | 16 | 1 | 1 | 3 | 1 | 31.2 | 6.4 | 8.4 | 28.9 | 76 | 1.5 | 0.06 | 0.10 | 0.22 | 0.41 | 99.21 |
| 127 | 2 | 16 | 12 | 0 | 2 | 3 | 1 | 31 | 6.5 | 8.4 | 28.9 | 71.6 | 3 | 0.04 | 0.19 | 0.51 | 2.17 | 97.09 |
| 128 | 2 | 10 | 14 | 0 | 1 | 2 | 1 | 30.9 | 6.8 | 8.4 | 28.9 | 68.8 | 2 | 0.11 | 0.04 | 0.36 | 0.76 | 98.73 |
| 129 | 3 | 15 | 11 | 0 | 4 | 1 | 0 | 30.7 | 6.6 | 8.4 | 28.9 | 77.2 | 4.1 | 0.08 | 0.44 | 1.42 | 0.87 | 97.19 |
| 130 | 1 | 33 | 15 | 0 | 3 | 2 | 1 | 30.7 | 6.4 | 8.3 | 28.9 | 61.6 | 4 | 0.18 | 0.21 | 0.33 | 0.90 | 98.37 |
| 131 | 4 | 5 | 22 | 0 | 1 | 1 | 2 | 30.7 | 6.6 | 8.2 | 24.6 | 110.8 | 3.8 | 0.10 | 0.12 | 0.65 | 0.97 | 98.15 |
| 132 | 2 | 8 | 21 | 0 | 1 | 2 | 1 | 30.7 | 6.6 | 8.18 | 28.7 | 86 | 2 | 0.04 | 0.09 | 0.41 | 0.66 | 98.80 |
| 133 | 1 | 18 | 18 | 0 | 1 | 2 | 1 | 30.7 | 6.8 | 8.3 | 28.8 | 128.8 | 2.2 | 0.05 | 0.09 | 0.32 | 1.01 | 98.53 |
| 134 | 1 | 24 | 32 | 0 | 1 | 1 | 0 | 30.5 | 6.4 | 8.4 | 28.43 | 95.2 | 2 | 0.09 | 0.14 | 0.40 | 1.04 | 98.33 |
| 135 | 1 | 33 | 28 | 1 | 1 | 1 | 0 | 30.5 | 6.4 | 8.4 | 28.7 | 101.6 | 2.5 | 0.04 | 0.06 | 0.74 | 1.18 | 97.98 |
| 136 | 2 | 24 | 43 | 0 | 1 | 1 | 1 | 30.5 | 6.3 | 8.3 | 28.7 | 75.6 | 2.1 | 0.05 | 0.07 | 0.44 | 1.07 | 98.36 |

Output ($O^P$): Polychaeta, Output ($O^B$): Bivalva, Output ($O^G$): Gastropoda, Output ($O^S$): Scaphopoda, Output ($O^A$): Amphipod, Output ($O^C$): Cumacea, Output ($O^E$): Echinodermata

Input 1: Temperature, Input 2: Dissolved Oxygen, Input 3: pH, Input 4: Salinity, Input 5: Total Suspended Solid (TSS)

Input 6: Organic matter, Input 7: Medium sand (%) 425 μm, Input 8: Fine sand (%) 250 μm, Input 9: Very fine sand (%) 125 μm, Input 10: Silt (%) 63 μm, Input 11: Silt and clay (%) 63≥ μm.

CHAPIN, F. S., CHAPIN, C., MATSON, P. A. & VITOUSEK, P. 2011. *Principles of Terrestrial Ecosystem Ecology*, Springer New York.
CUFF, D. & GOUDIE, A. 2009. *The Oxford Companion to Global Change*, Oxford University Press, USA.
HILLEWAERT, H. 2006. Scheme of eutrophication. *In:* EUTROPHICATION.JPG (ed.).
HOSSEINZADEH LOTFI, F., JAHANSHAHLOO, G. R., KHODABAKHSHI, M., ROSTAMY-MALKHLIFEH, M., MOGHADDAS, Z. & VAEZ-GHASEMI, M. 2013. A Review of Ranking Models in Data Envelopment Analysis. *Journal of Applied Mathematics,* 2013**,** 20.
IPCC 2014. *Climate Change 2014: Impacts, Adaptation, and Vulnerability. Part A: Global and Sectoral Aspects. Contribution of Working Group II to the Fifth Assessment Report of the Intergovernmental Panel on Climate Change [Field, C.B., V.R. Barros, D.J. Dokken, K.J. Mach, M.D. Mastrandrea, T.E. Bilir, M. Chatterjee, K.L. Ebi, Y.O. Estrada, R.C. Genova, B. Girma, E.S. Kissel, A.N. Levy, S. MacCracken, P.R. Mastrandrea, and L.L. White (eds.)],* Cambridge, United Kingdom and New York, NY, USA, Cambridge University Press.

**Table1:** Main attribute of the four sampling locations

| Sampling Station | Condition | Transects (Distance from the shore) | | Coordinate |
|---|---|---|---|---|
| **Teluk Bahang** | Fishery village, less human activities relatively pollution | 200 m | TB1 | N 05'46.28" , E 100'20.74" |
| | | 400 m | TB2 | N 05'46.62" , E 100'20.81" |
| | | 600 m | TB3 | N 05'47.09" , E 100'20.9" |
| | | 800 m | TB4 | N 05'47.51" , E 100'20.95" |
| | | 1000 m | TB5 | N 05'47.95" , E 100'21.02" |
| | | 1200 m | TB6 | N 05'48.39" , E 100'20.96" |
| **Teluk Aling** | less human activities, near the Center for Marine and Coastal Studies (USM) | 200 m | TA1 | N 05'46.9"  , E 100'20.04" |
| | | 400 m | TA2 | N 05'46.91" , E 100'19.99" |
| | | 600 m | TA3 | N 05'47.73" , E 100'19.95" |
| | | 800 m | TA4 | N 05'48.16" , E 100'19.94" |
| | | 1000 m | TA5 | N 05'48.65" , E 100'19.99" |
| | | 1200 m | TA6 | N 05'49.11" , E 100'20.04" |
| **Teluk Ketapag** | Pristine (control site) | 200 m | TK1 | N 05'46.5"  , E 100'17.86" |
| | | 400 m | TK2 | N 05'46.45" , E 100'17.39" |
| | | 600 m | TK3 | N 05'46.51" , E 100'16.96" |
| | | 800 m | TK4 | N 05'46.65" , E 100'16.53" |
| | | 1000 m | TK5 | N 05'46.79" , E 100'16.11" |
| | | 1200 m | TK6 | N 05'46.86" , E 100'15.71" |
| **Pantai Acheh** | Near the mangrove, highest turbidity | 200 m | PA1 | N 05'39.42" , E 100'16.7" |
| | | 400 m | PA2 | N 05'39.49" , E 100'16.22" |
| | | 600 m | PA3 | N 05'39.51" , E 100'15.81" |
| | | 800 m | PA4 | N 05'39.5"  , E 100'15.41" |
| | | 1000 m | PA5 | N 05'39.62" , E 100'15.02" |
| | | 1200 m | PA6 | N 05'39.83" , E 100'14.72" |

**Table2:** DMUs with eleven inputs and a single output associated with each of the seven species of macrobenthos

| | Inputs | | | | | | | | | | | Outputs | | | | | | |
|---|---|---|---|---|---|---|---|---|---|---|---|---|---|---|---|---|---|---|
| | Tem ($I_1$) | DO ($I_2$) | pH ($I_3$) | Salinity ($I_4$) | TSS ($I_5$) | O.M ($I_6$) | M.S ($I_7$) | F.S ($I_8$) | VF.S ($I_9$) | S ($I_{10}$) | SC ($I_{11}$) | Polychaeta ($O^P$) | Bivalva ($O^B$) | Gastropoda, ($O^G$) | Scaphopod a ($O^S$) | Amphipod ($O^A$) | Cumacea ($O^C$) | Echinodermata ($O^E$) |
| 1 | 31.2 | 5.15 | 8.2 | 30 | 100.4 | 3 | 6.40 | 3.74 | 2.72 | 2.29 | 81.58 | 2 | 48 | 12 | 0 | 1 | 3 | 1 |
| 2 | 31 | 5.6 | 8.2 | 30 | 120 | 3.5 | 0.33 | 0.47 | 0.93 | 1.50 | 96.77 | 2 | 3 | 3 | 4 | 1 | 3 | 0 |
| 3 | 31.5 | 5.58 | 8.3 | 30 | 77.2 | 2.9 | 0.18 | 0.15 | 0.78 | 1.47 | 97.42 | 1 | 4 | 18 | 0 | 1 | 5 | 1 |
| .* | . | . | . | . | . | . | . | . | . | . | . | . | . | . | . | . | . | . |
| . | . | . | . | . | . | . | . | . | . | . | . | . | . | . | . | . | . | . |
| . | . | . | . | . | . | . | . | . | . | . | . | . | . | . | . | . | . | . |
| 134 | 30.5 | 6.6 | 8.4 | 28.43 | 95.2 | 2 | 0.09 | 0.14 | 0.40 | 1.04 | 98.33 | 1 | 24 | 32 | 0 | 1 | 1 | 0 |
| 135 | 30.5 | 6.4 | 8.4 | 28.7 | 101.6 | 2.5 | 0.04 | 0.06 | 0.74 | 1.18 | 97.98 | 1 | 33 | 28 | 1 | 1 | 1 | 0 |
| 136 | 30.5 | 6.3 | 8.3 | 28.7 | 75.6 | 2.1 | 0.05 | 0.07 | 0.44 | 1.07 | 98.36 | 2 | 24 | 43 | 0 | 1 | 1 | 1 |

*The full list of data is available in the Appendix.

Output ($O^P$): Polychaeta, Output ($O^B$): Bivalva, Output ($O^G$): Gastropoda, Output ($O^S$): Scaphopoda, Output ($O^A$): Amphipod, Output ($O^C$): Cumacea, Output ($O^E$): Echinodermata

Input 1: Temperature, Input 2: Dissolved Oxygen, Input 3: pH, Input 4: Salinity, Input 5: Total Suspended Solid (TSS)

Input 6: Organic matter, Input 7: Medium sand (%) 425 μm, Input 8: Fine sand (%) 250 μm, Input 9: Very fine sand (%) 125 μm, Input 10: Silt (%) 63 μm, Input 11: Silt and clay (%) 63≥ μm.

**Table3:** Eigenvalues of Principal Components Analysis

| | Eigenvalues | % Total variance | Cumulative eigenvalue | Cumulative % |
|---|---|---|---|---|
| 1 | 5.989617 | 29.94809 | 5.98962 | 29.94809 |
| 2 | 2.550573 | 12.75287 | 8.54019 | 42.70095 |
| 3 | 1.764573 | 8.82286 | 10.30476 | 51.52382 |

**Table4:** Rotated principal component loadings for 20 standardized sediment parameters and environmental factors. The three PCA factors had eigenvalues more than 1.

| Variables | PC 1 | PC 2 | PC 3 |
|---|---|---|---|
| Transect | -0.23 | 0.101 | 0.259 |
| Temperature (°C) | -0.045 | 0.332 | -0.184 |
| Dissolved oxygen (mg/L) | -0.119 | -0.471 | -0.004 |
| pH | -0.107 | 0.149 | -0.258 |
| Salinity (ppt) | 0.012 | 0.311 | 0.004 |
| Total Suspended Solid (mg/L) | -0.097 | -0.349 | -0.028 |
| Organic matter (%) | 0.387 | -0.121 | -0.122 |
| Coarse Sand | 0.371 | -0.052 | 0.058 |
| Medium sand | 0.387 | -0.038 | 0.048 |
| Fine sand | 0.389 | -0.036 | 0.053 |
| Very fine sand | 0.369 | 0.019 | 0.005 |
| Silt | 0.216 | 0.096 | -0.055 |
| Silt and clay | -0.401 | 0.013 | -0.033 |

**Table5:** The Spearman Rank Correlation of macrobenthos and three principal components at all locations.

| | Mean | Std. Dev. | PC1 | PC 2 | PC 3 |
|---|---|---|---|---|---|
| Total Abundance | 2901.91 | 1325.92 | 0.73 | 0.05 | 0.09 |
| Diversity | 2.184 | 0.38 | 0.19 | -0.08 | -0.03 |
| Mollusca | 1968.21 | 1019.50 | 0.69 | -0.03 | 0.16 |
| Polychaeta | 413.52 | 341.25 | 0.66 | -0.01 | -0.02 |
| Crustacean | 459.27 | 117.28 | 0.36 | 0.27 | -0.13 |
| Echinodermata | 60.95 | 35.1 | 0.60 | -0.02 | 0.01 |

**Table6:** Results of model (5) of step 6

| | Tem | DO | pH | Salinity | TSS | O.M | M.S | F.S | VF.S | S | SC |
|---|---|---|---|---|---|---|---|---|---|---|---|
| **Polychaeta** | 0.028 | 0.250 | 0.029 | 0.031 | 0.361 | 0.484 | 1.164 | 0.859 | 0.732 | 0.632 | 0.030 |
| **Bivalva** | 0.034 | 0.250 | 0.024 | 0.034 | 0.350 | 0.486 | 1.165 | 0.858 | 0.730 | 0.625 | 0.039 |
| **Gastropoda** | 0.062 | 0.272 | 0.029 | 0.048 | 0.376 | 0.501 | 1.165 | 0.858 | 0.737 | 0.665 | 0.050 |
| **Scaphopoda** | 0.048 | 0.261 | 0.028 | 0.028 | 0.357 | 0.465 | 1.160 | 0.862 | 0.736 | 0.625 | 0.040 |
| **Amphipod** | 0.024 | 0.241 | 0.027 | 0.024 | 0.343 | 0.488 | 1.167 | 0.857 | 0.733 | 0.630 | 0.032 |
| **Cumacea** | 0.027 | 0.250 | 0.025 | 0.024 | 0.358 | 0.501 | 1.169 | 0.856 | 0.726 | 0.628 | 0.042 |
| **Echinodermata** | 0.114 | 0.293 | 0.079 | 0.088 | 0.375 | 0.547 | 1.172 | 0.853 | 0.714 | 0.638 | 0.106 |

Tem: Temperature, DO: Dissolved Oxygen, TSS: Total Suspended Solid (TSS) , O.M: Organic matter, M.S: Medium sand (%) 425 μm, F.S: Fine sand (%) 250 μm, VF.S: Very fine sand (%) 125 μm, S: Silt (%) 63 μm, SC: Silt and clay (%) 63≥ μm

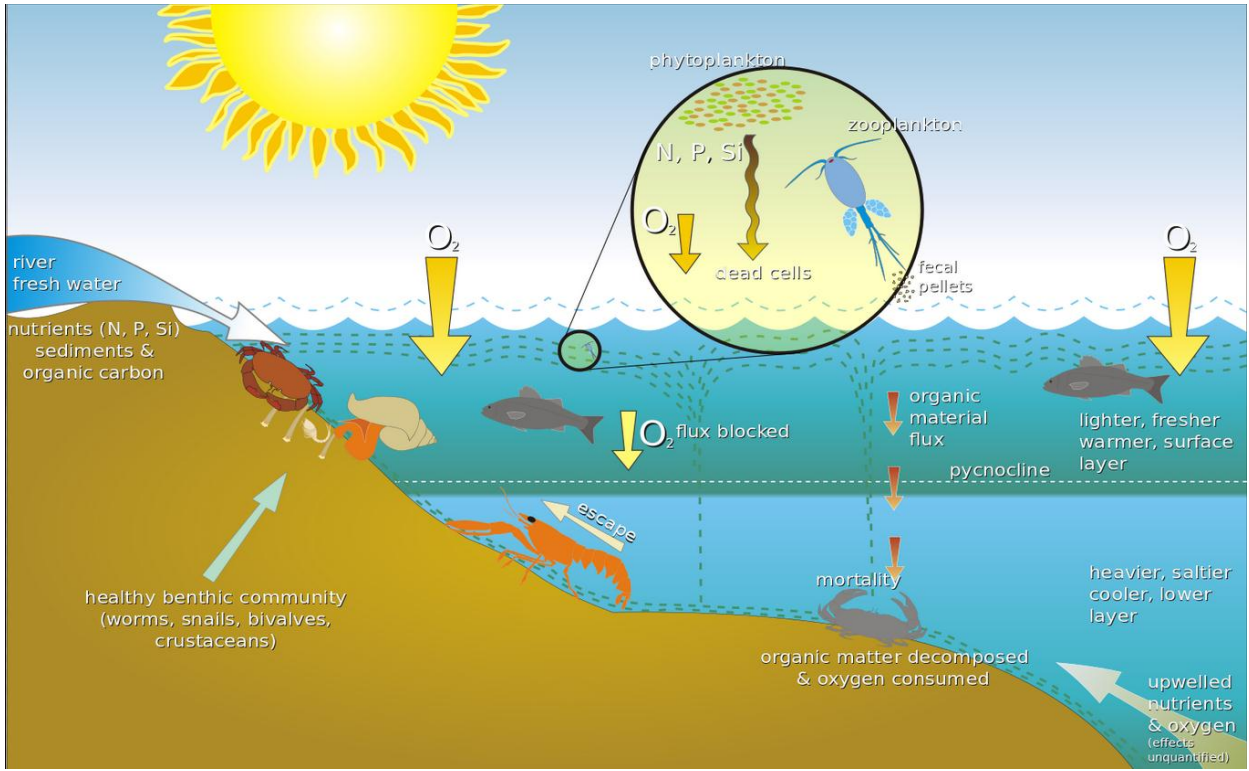

**Figure 1:** Illustration of Eutrophication (Image: (Hillewaert, 2006))

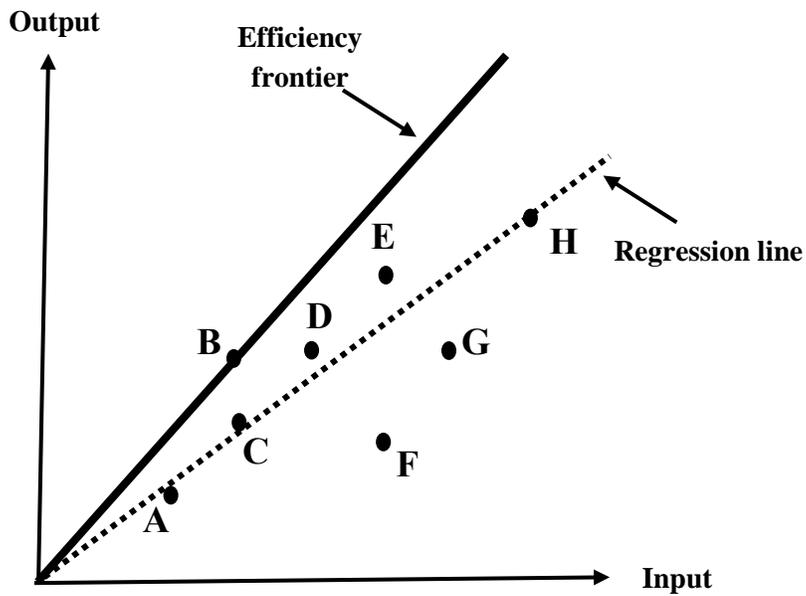

**Figure 2:** Comparison between efficiency frontier and regression line

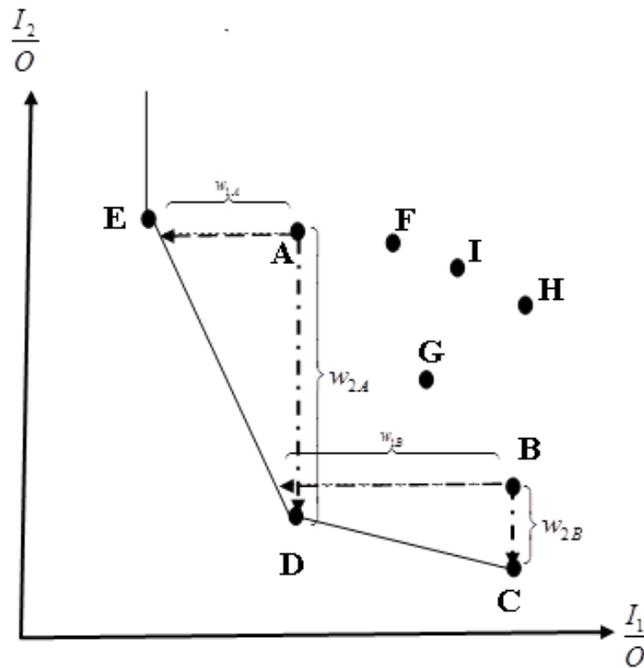

**Figure 3:** Measuring the distance of the inputs from the efficiency frontier in Stage 1

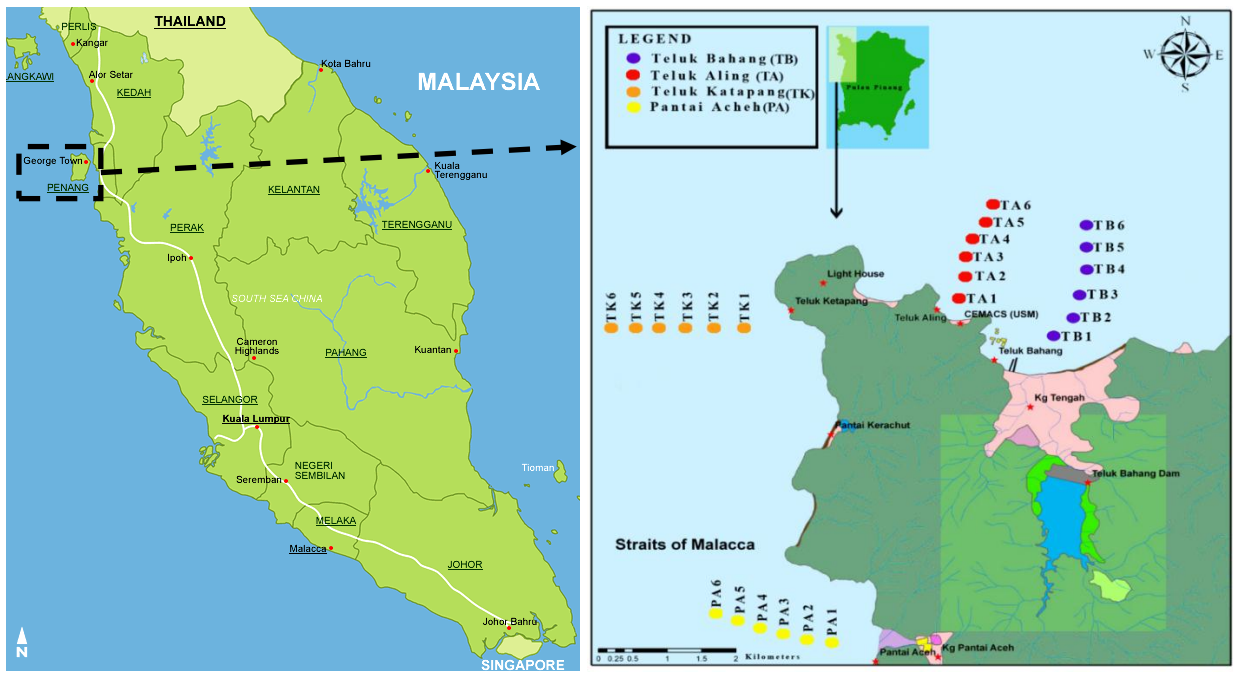

**Figure 4:** Location of macrobenthic sampling stations (Penang National Park) in the coastal waters of Straits of Malacca. Transect (1=200 m, 2=400 m, 3=600 m, 4=800 m, 5=1000 m and 6=1200 m).

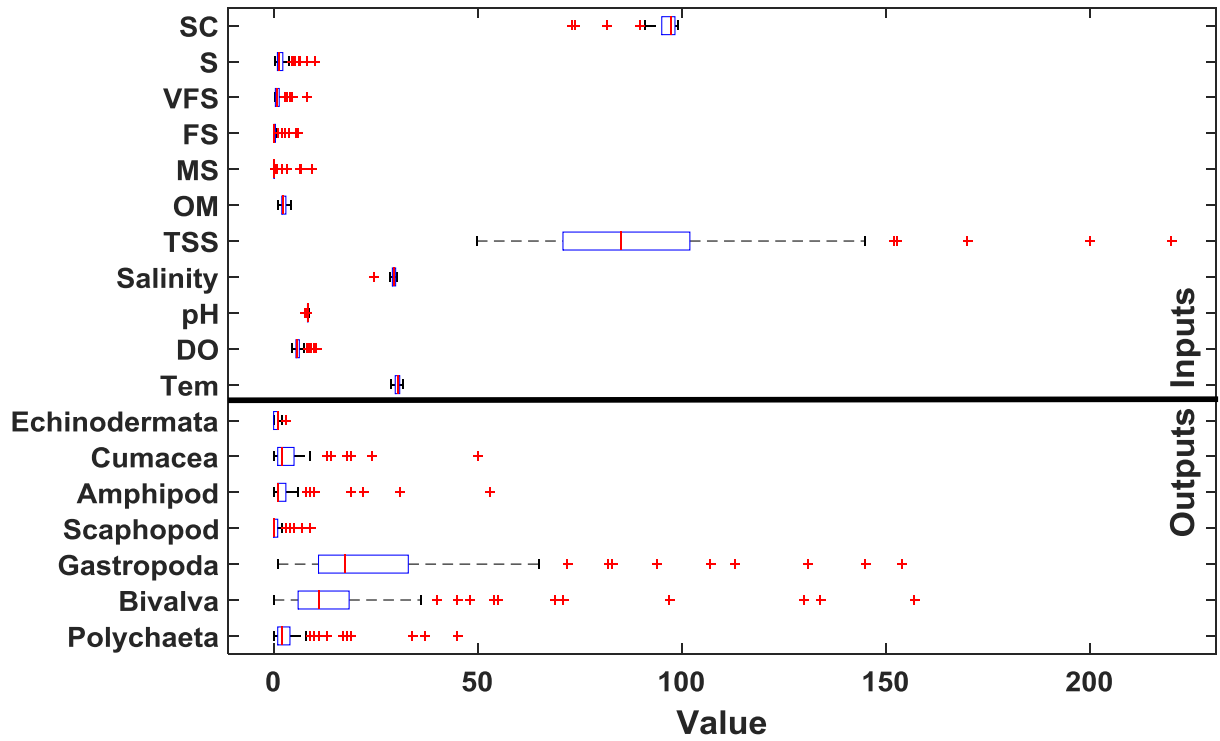

**Figure 5:** Distribution of the input and output variables based on sampled data

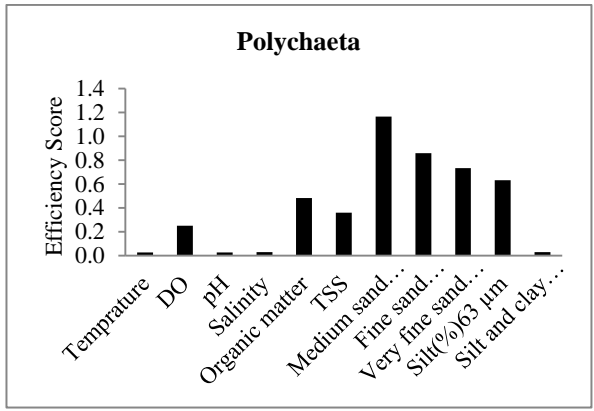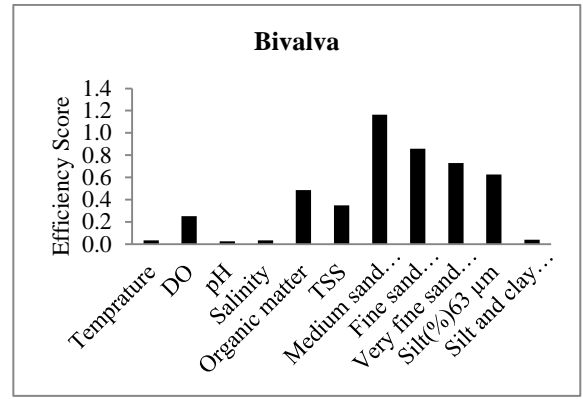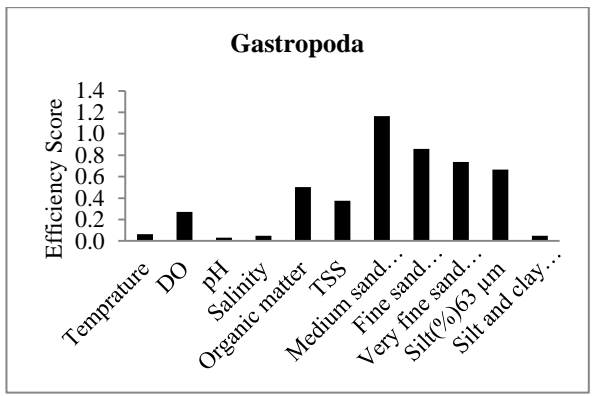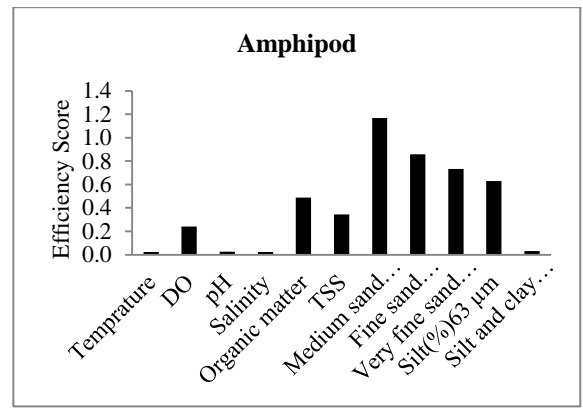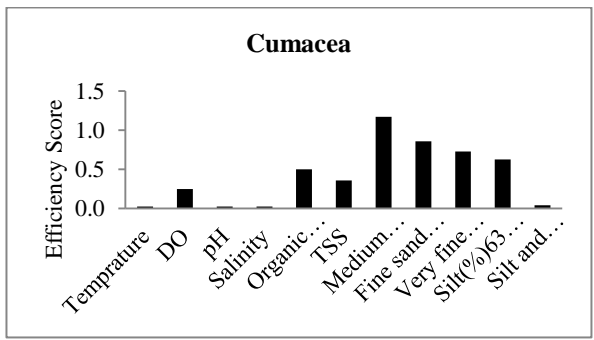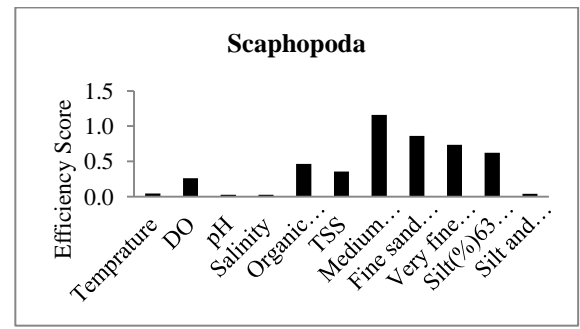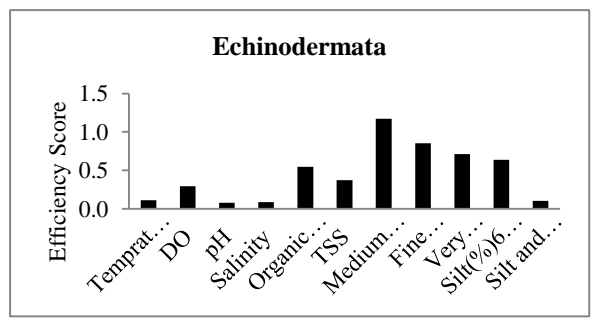

**Figure 6:** Efficiency score of environmental parameters for all outputs of study area